\documentstyle[multicol,aps,epsf]{revtex}
\begin{document}
\title{Connectivity of Growing Random Networks}
\author{P.~L.~Krapivsky$^{1,2}$, S.~Redner$^1$, and F. Leyvraz$^3$}
\address{$^1$Center for BioDynamics, Center for Polymer Studies, 
and Department of Physics, Boston University, Boston, MA, 02215}
\address{$^2$CNRS, IRSAMC, Laboratoire de Physique Quantique,
Universite' Paul Sabatier, 31062 Toulouse, France}
\address{$^3$Centro Internacional de Ciencias, Cuernavaca, Morelos, Mexico}

\maketitle
\begin{abstract}
  A solution for the time- and age-dependent connectivity distribution of a
  growing random network is presented.  The network is built by adding sites
  which link to earlier sites with a probability $A_k$ which depends on the
  number of pre-existing links $k$ to that site.  For homogeneous connection
  kernels, $A_k\sim k^\gamma$, different behaviors arise for $\gamma<1$,
  $\gamma>1$, and $\gamma=1$.  For $\gamma<1$, the number of sites with $k$
  links, $N_k$, varies as stretched exponential.  For $\gamma>1$, a single
  site connects to nearly all other sites.  In the borderline case $A_k\sim
  k$, the power law $N_k\sim k^{-\nu}$ is found, where the exponent $\nu$ can
  be tuned to any value in the range $2<\nu<\infty$.

\smallskip\noindent{PACS numbers: 02.50.Cw, 05.40.-a, 05.50.+q, 87.18.Sn}
\end{abstract}
\begin{multicols}{2}
  
\noindent
Random networks play an important role in epidemiology, ecology (food webs),
and many other fields.  The geometry of such fixed topology networks have
been extensively investigated \cite{bol,jan,kauf,wasserman,der,flyv,parisi}.
However, networks based on human interactions, such as transportation
systems, electrical distribution systems, biological systems, and the
Internet are open and continuously growing and new approaches are rapidly
developing to understand their structure and time evolution
\cite{sole,jain,klein,kum,brod}.

In this Letter, we apply a rate equation approach to solve the growing random
network (GRN) model, a special case of which was introduced in
\cite{barabasi} to account for the distribution of citations and other
growing networks\cite{barabasi,lotka,sor,redner,jose,huberman}.  Our approach
is ideally-suited for the GRN and is much simpler than the standard
probabilistic\cite{bol} or generating function\cite{jan} techniques.  The
rate equation formulation can be adapted to study more general evolving graph
systems, such as networks with site deletion and link re-arrangement.

\begin{figure}
  \narrowtext \epsfxsize=1.8in \hskip 0.6in
\epsfbox{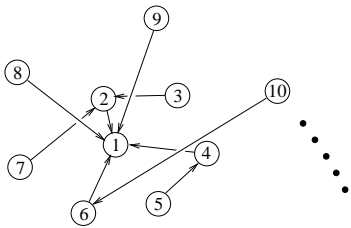} \vskip 0.1in
\caption{Schematic illustration of the evolution of the growing random network.
  Sites are added sequentially and a single link joins the new site to
  an earlier site.
\label{network}}
\end{figure}
  
The GRN model is defined as follows.  At each time step, a new site is added
and a directed link to one of the earlier sites is created.  In terms of
citations, we may interpret the sites in Fig.~\ref{network} as publications,
and the directed link from one paper to another as a citation to the earlier
publication.  This growing network has a directed tree graph topology where
the basic elements are sites which are connected by directed links.  The
structure of this graph is determined by the connection kernel $A_k$, which
is the probability that a newly-introduced site links to an existing site
with $k$ links ($k-1$ incoming and 1 outgoing).  We will solve for the
connectivity distribution $N_k(t)$, defined as the average number of sites
with $k$ links as a function of the connection kernel $A_k$.

We focus on a class of homogeneous connection kernels, $A_k=k^\gamma$, with
$\gamma\geq 0$ reflecting the tendency of preferential linking to popular
sites.  As we shall show, the connectivity distribution crucially depends on
whether $\gamma$ smaller than, larger than, or equal to unity.  For
$\gamma<1$, the connectivity distribution decreases as a stretched
exponential in $k$.  The case $\gamma>1$ leads to phenomenon akin to
gelation\cite{agg} in which a single ``gel'' site connects to nearly every
other site of the graph.  For $\gamma>2$, this phenomenon is so extreme that
the number of connections between other sites is finite in an infinite graph.
A power law distribution $N_k\sim k^{-\nu}$ arises {\em only} for $\gamma=1$.
In this case, finer details of the dependence of the connection kernel on $k$
affect the exponent $\nu$.  Hence we consider a more general class of {\em
  asymptotically} linear connection kernels, $A_k\sim k$ as $k\to\infty$.  We
show that $\nu$ is tunable to any value in the range $2<\nu<\infty$.  In
particular, we can naturally generate values of $\nu$ between 2 and 3, as
observed in the web graph\cite{klein,kum,brod} and in movie actor
collaboration networks\cite{barabasi}.

The rate equations for the time evolution of the connectivity
distribution $N_k(t)$ are
\begin{equation}
\label{Nk}
{d N_k\over dt}={1\over M_\gamma}
\left[(k-1)^\gamma N_{k-1}-k^\gamma N_k\right]+\delta_{k1}.
\end{equation}
The first term accounts for the process in which a site with $k-1$ links is
connected to the new site, leading to a gain in the number of sites with $k$
links.  This happens with probability $(k-1)^\gamma/M_\gamma$, where
$M_\gamma(t)=\sum j^\gamma N_j(t)$ provides the proper normalization.  A
corresponding role is played by the second (loss) term on the right-hand side
of Eq.~(\ref{Nk}).  The last term accounts for the continuous introduction of
new sites with no incoming links.

We start by finding the low-order moments $M_n(t)$ of the connectivity
distribution.  Summing Eqs.~(\ref{Nk}) over all $k$ gives the rate
equation for the total number of sites, $\dot M_0=1$, whose solution is
$M_0(t)= M_0(0)+t$.  The first moment (the total number of bond
endpoints) obeys $\dot M_1=2$, which gives $M_1(t)= M_1(0)+2t$.  The
first two moments are therefore {\em independent\/} of $\gamma$, while
higher moments and the connectivity distribution itself do depend on
$\gamma$.

For the linear connection kernel, Eqs.~(\ref{Nk}) can be solved for an
arbitrary initial condition.  We limit ourselves to the most
interesting asymptotic regime ($t\to\infty$) where the initial condition
is irrelevant.  Using $M_1=2t$, we solve the first few of
Eqs.~(\ref{Nk}) and obtain $N_1=2t/3$, $N_2=t/6$, {\it etc.}, which
implies that the $N_k$ grow linearly with time.  Accordingly, we
substitute $N_k(t)=t\,n_k$ in Eqs.~(\ref{Nk}) to yield the recursion
relation $n_k=n_{k-1} (k-1)/(k+2)$.  Solving for $n_k$ then gives
\begin{equation}
\label{nk1}
n_k={4\over k(k+1)(k+2)}. 
\end{equation}

To solve the model with a sub-linear connection kernel, $0<\gamma<1$,
notice that $M_\gamma$ satisfies the obvious inequalities $M_0\leq
M_\gamma\leq M_1$.  Consequently, in the long-time limit
\begin{equation}
\label{mg}
M_\gamma=\mu t, \qquad  1\leq \mu \leq 2,
\end{equation}
with a yet undetermined prefactor $\mu=\mu(\gamma)$.  Now substituting
$N_k(t)=t\,n_k$ and $M_\gamma=\mu t$ into Eqs.~(\ref{Nk}) and again solving
for $n_k$ we obtain
\begin{equation}
\label{Nkg}
n_k={\mu\over k^{\gamma}}\prod_{j=1}^{k}
\left(1+{\mu\over j^\gamma}\right)^{-1},
\end{equation}
whose asymptotic behavior is
\begin{eqnarray}
\label{cases} 
n_k\sim\cases{
k^{-\gamma}\exp
\left[-\mu\left({{k^{1-\gamma}-2^{1-\gamma}}\over 1-\gamma}\right)\right]
&${1\over 2}<\gamma<1$,\cr
&\cr
k^{\mu^2-1\over 2}\exp\left[-2\mu\,\sqrt{k} \right] 
&$\gamma={1\over 2}$,\cr
&\cr
k^{-\gamma}\exp\left[-\mu\, {k^{1-\gamma}\over 1-\gamma}
+{\mu^2\over 2}\, {k^{1-2\gamma}\over 1-2\gamma}\right]
&${1\over 3}<\gamma<{1\over 2}$,\cr}
\end{eqnarray}
{\it etc}.  This pattern in (\ref{cases}) continues {\it ad infinitum}:
Whenever $\gamma$ decreases below $1/m$, with $m$ a positive integer, an
additional term in the exponential arises from the now relevant contribution
of the next higher-order term in the expansion of the product in
Eq.~(\ref{Nkg}).

\begin{figure}
  \narrowtext \epsfxsize=2.0in \hskip 0.3in
\epsfbox{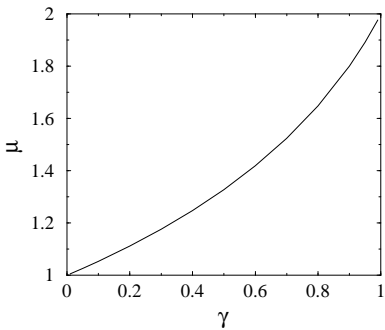} \vskip 0.10in
\caption{The amplitude $\mu$ in $M_\gamma(t)=\mu t$ versus $\gamma$.
\label{mu-vs-gamma}}
\end{figure}

To complete the solution for the $n_k$, we need to establish the dependence
of the amplitude $\mu$ on $\gamma$.  Using the defining relation
$M_\gamma/t=\mu=\sum_{k\geq 1}k^\gamma n_k$, together with Eq.~(\ref{Nkg}),
we obtain the implicit relation for $\mu(\gamma)$
\begin{equation}
\label{mu}
\mu=\sum_{k=2}^\infty \prod_{j=2}^{k}
\left(1+{\mu\over j^\gamma}\right)^{-1}.
\end{equation}
Despite the simplicity of this exact expression, it is not easy to
extract explicit information except for the limiting cases $\gamma=0$
and $\gamma=1$, where $\mu=1$ and $\mu=2$ respectively, and the
corresponding connectivity distributions are given by $n_k=2^{-k}$ and
by Eq.~(\ref{nk1}).  However, numerical evaluation shows that $\mu$
varies smoothly between 1 and 2 as $\gamma$ increases from 0 to 1
(Fig.~\ref{mu-vs-gamma}).  This result, together with Eq.~(\ref{Nkg}),
provides a comprehensive description of the connectivity distribution in
the regime $0\leq \gamma\leq 1$.  It is worth emphasizing that for
$0.8\alt \gamma\alt 1$, $n_k$ depends weakly on $\gamma$ for $1\leq
k\leq 1000$.  Thus, it is difficult to discriminate between different
$\gamma$'s and even to distinguish a power law from a stretched
exponential in the GRN model.  This subtlety was already encountered in
the analysis of the citation distribution\cite{sor,redner}.

A striking feature of the GRN model is that we can ``tune'' the exponent
$\nu$ by augmenting the linear connection kernel to the asymptotically
linear connection kernel, with $A_k\to a_\infty k$ as $k\to\infty$, but
otherwise {\em arbitrary}.  For this asymptotically linear kernel, by
repeating the steps leading to Eq.~(\ref{Nkg}) we find
\begin{equation}
\label{nkgen}
n_k={\mu\over A_k}\,
\prod_{j=1}^k \left(1+{\mu\over A_j}\right)^{-1}.
\end{equation}
Expanding the product in Eq.~(\ref{nkgen}) leads to $n_k\sim k^{-\nu}$ with
$\nu=1+\mu/a_\infty$, while the amplitude $\mu$ is found from
\begin{equation}
\label{mugen}
\mu=A_1\sum_{k=2}^\infty\prod_{j=2}^k \left(1+{\mu\over A_j}\right)^{-1}.
\end{equation}
As an explicit example, consider the connection kernel $A_1=1$ and
$A_k=a_\infty k$ for $k\geq 2$.  In this case, we can reduce
Eq.~(\ref{mugen}) to a quadratic equation from which we obtain
$\nu=(3+\sqrt{1+8/a_\infty})/2$ which can indeed be tuned to {\em any} value
larger than 2.

The GRN model with super-linear connection kernels, $\gamma>1$, exhibits
a ``winner take all'' phenomenon, namely the emergence of a single
dominant ``gel'' site which is linked to almost every other site.  A
particularly singular behavior occurs for $\gamma>2$, where there is a
non-zero probability that the initial site is connected to every other
site of the graph.  To determine this probability, it is convenient to
consider a discrete time version process where one site is introduced at
each step which always links to the initial site.  After $N$ steps, the
probability that the new site will link to the initial site is
$N^\gamma/(N+N^\gamma)$.  This pattern continues indefinitely with
probability
\begin{equation}
\label{prob}
{\cal P}=\prod_{N=1}^\infty {1\over 1+N^{1-\gamma}}.
\end{equation}
Clearly, ${\cal P}=0$ when $\gamma\leq 2$ but ${\cal P}>0$ when $\gamma>2$.
Thus for $\gamma>2$ there is a non-zero probability that the initial site
connects to all other sites.  

To determine the behavior for general $\gamma>1$, we need the asymptotic time
dependence of $M_\gamma$.  To this end, it is useful to consider the
discretized version of the master equations Eq.~(\ref{Nk}), where the time
$t$ is limited to integer values.  Then $N_k(t)=0$ whenever $k>t$ and the
rate equation for $N_k(k)$ immediately leads to
\begin{eqnarray}
N_k(k) &=& \frac{(k-1)^\gamma N_{k-1}(k-1)}{M_\gamma(k-1)}\nonumber \\
       &=& N_2(2)\prod_{j=2}^{k-1}\frac{j^\gamma}{M_\gamma(j)}.
\label{eq:Nkk}
\end{eqnarray}
{}From this and the obvious fact that $N_k(k)$ must be less than unity, it
follows that $M_\gamma(t)$ cannot grow more slowly than $t^\gamma$.  On the
other hand, $M_\gamma(t)$ cannot grow faster than $t^\gamma$ as follows
from the estimate
\begin{eqnarray}
M_\gamma(t)&=&\sum_{k=1}^t k^\gamma N_k(t)\nonumber \\
  &\leq& t^{\gamma-1}\sum_{k=1}^t k N_k(t)=t^{\gamma-1}M_1(t)
\end{eqnarray}
Thus $M_\gamma\propto t^\gamma$.  In fact, the amplitude of $t^\gamma$ is
unity as will be derived self-consistently after solving for the $N_k$'s.

We now use $M_\gamma\sim t^\gamma$ in the rate equations to solve recursively
for each $N_k$.  Starting with the equation $\dot N_1=1-N_1/M_\gamma$, the
second term on the right-hand side is sub-dominant; neglecting this term gives
$N_1=t$.  Continuing this same line of reasoning for each successive rate
equation gives the leading behavior of $N_k$,
\begin{equation}
\label{Nkgg}
N_k = J_kt^{k-(k-1)\gamma}\quad {\rm for}\quad k\geq 1,
\end{equation}
with $J_k=\prod_{j=1}^{k-1}j^\gamma/[1+j(1-\gamma)]$.  This pattern of
behavior for $N_k$ continues as long as its exponent $k-(k-1)\gamma$ remains
positive, or $k<\gamma/(\gamma-1)$.  The full behavior of the $N_k$ may be
determined straightforwardly by keeping the next correction terms in the rate
equations.  For example, $N_1=t-t^{2-\gamma}/(2-\gamma)+\ldots$.

For $k>\gamma/(\gamma-1)$, each $N_k$ has a finite limiting value in the
long-time limit.  Since the total number of connections equals $2t$ and $t$
of them are associated with $N_1$, the remaining $t$ links must all connect
to a single site which has $t$ connections (up to corrections which grow no
faster than sub-linearly with time).  Consequently the amplitude of $M_\gamma$
equals unity, as argued above.

Thus for super-linear kernels, the GRN undergoes an infinite sequence of
connectivity transitions as a function of $\gamma$.  For $\gamma>2$ all but a
finite number of sites are linked to the ``gel'' site which has the rest of
the links of the network.  This is the ``winner take all'' situation.  For
$3/2<\gamma<2$, the number of sites with two links grows as $t^{2-\gamma}$,
while the number of sites with more than two links is again finite.  For
$4/3<\gamma<3/2$, the number of sites with three links grows as
$t^{3-2\gamma}$ and the number with more than three is finite.  Generally for
${m+1\over m}<\gamma<{m\over m-1}$, the number of sites with more than $m$
links is finite, while $N_k\sim t^{k-(k-1)\gamma}$ for $k\leq m$.
Logarithmic corrections also arise at the transition points.

The connectivity distribution leads to an amusing consequence for the most
popular site.  Its connectivity $k_{\rm max}$ is determined by
$\sum_{k>k_{\rm max}} N_k=1$, that is, there is one site whose
connectivity lies in the range $(k_{\rm max},\infty)$.  This criterion
gives
\begin{equation}
\label{kmax}
k_{\rm max}\sim \cases{
            (\ln t)^{1/(1-\gamma)}  & $0\leq \gamma<1$; \cr
            t^{1/(\nu-1)}           & asymptotically linear; \cr  
            t                       & super-linear.}
\end{equation}
Since $t$ also equals the total number of sites, we can compare this
prediction about the most popular site with available data from the Institute
of Scientific Information based on 783,339 papers with 6,716,198 total
citations (details in Ref.~\cite{redner}).  Here the most cited paper had
8,904 citations.  This accords with the first line of Eq.~(\ref{kmax}) for
$\gamma\approx 0.86$, and also with the second when $\nu\approx 2.5$.

In addition to the connectivity of a site, we also may ask about its {\em
  age}.  Within the GRN model, older sites should clearly be more highly
connected.  We quantify this feature and also determine how the connection
kernel affects the combined age and connectivity distribution.  Note that our
model does {\em not\/} have explicit aging where the connection kernel
depends on the age of each site; this feature is treated in Ref.~\cite{jose}.

Let $c_k(t,a)$ be the average number of sites of age $a$ which have $k-1$
incoming links at time $t$.  Here age $a$ means that the site was introduced
at time $t-a$.  The quantity $c_k(t,a)$ evolves according to
\begin{equation}
\label{ck}
{\partial c_k \over \partial t}+{\partial c_k \over \partial a}
={1\over M_\gamma}\left[(k-1)^\gamma c_{k-1}-k^\gamma c_k\right]
+\delta_{k1}\delta(a).
\end{equation}
The second term on the left-hand side accounts for the aging of sites, while
the right-hand side accounts for the (age independent) connection changing
processes.  Consider first the linear kernel, $A_k=k$.  Let us focus again on
the most interesting limit, namely asymptotic behavior.  Then we can
disregard the initial condition and write $M_1(t)=2t$.  This transforms
Eqs.~(\ref{ck}) into
\begin{equation}
\label{ck1}
\left({\partial \over \partial t}+{\partial \over \partial a}\right)c_k 
={(k-1) c_{k-1}-k c_k\over 2t}+\delta_{k1}\delta(a).
\end{equation}
The homogeneous form of this equation suggests that solution should be
self-similar. Specifically, one can seek a solution as a function of the {\em
  single} variable $a/t$ rather than two separate variables,
$c_k(t,a)=f_k(a/t)$.  This simplifies the partial differential equation
(\ref{ck1}) into an ordinary differential equation for $f_k(x)$ which can be
easily solved.  In terms of the original variables of $a$ and $t$, we find
\begin{equation}
\label{ck1all}
c_k(t,a)=\sqrt{1-{a\over t}}\left[1-\sqrt{1-{a\over t}}\right]^{k-1}.
\end{equation}
Notice that this age distribution satisfies the normalization
requirement, $N_k(t)=\int_0^t da\,c_k(t,a)$.  As expected, young sites
(those with $a/t\to 0$) typically have a small connectivity while old
sites have large connectivity.  Further, old sites have a broad
distribution of connectivities up to a characteristic number which
asymptotically grows as $\langle k\rangle\sim (1-a/t)^{-1/2}$ as $a\to
t$.  These properties and related issues may be worthwhile to
investigate in citation and other information networks.

Similarly, we can obtain $c_k(t,a)$ for the GRN model with an arbitrary
homogeneous connection kernel\cite{kr} which grows slower than linearly in
$k$.  Assuming a self-similar solution $c_k(t,a)=f_k(a/t)$, applying a
Laplace transform, we find a recursion relation for $\hat f_k$ whose solution
is identical in structure to Eq.~(\ref{Nkg}).  Although it appears impossible
to perform the inverse Laplace transform in explicit form for arbitrary $k$,
we can compute $c_k(t,a)$ for small $k$; for example, we find
$c_1=(1-a/t)^{1/\mu}$.  The behavior also simplifies in the large-$k$ limit.
Here we find that the age of sites with $k$ links is peaked about the value
$a_k$ which satisfies
\begin{equation}
\label{akg}
{a_k\over t} \simeq \cases
{1-\exp\left(-\mu\,{k^{1-\gamma}\over 1-\gamma}\right) & $\gamma<1$;\cr
   1-{12\over{(k+3)(k+4)}}  & $\gamma=1$.}
\end{equation}
This shows how old sites are better connected.

In summary, we solved for both the connectivity distribution and the
age-dependent structure of the growing random network.  The most interesting
connectivity arises in a network with an asymptotically linear connection
kernel.  Here the number of sites with $k$ connections has the power-law form
$N_k\sim k^{-\nu}$, with $\nu$ tunable to any value in the range
$2<\nu<\infty$.  This accords with the connectivity distributions observed in
various contemporary examples of growing networks.

\medskip 

We are grateful to grants NSF INT9600232, NSF DMR9978902, and DGAPA IN112998
for financial support.  While writing this manuscript we learned of
Ref.\cite{doro} which overlaps some of our results.  We thank J. Mendes for
informing us of this work.

\end{multicols}
\end{document}